\documentstyle[twocolumn,prl,aps,epsfig]{revtex}
\newcommand{\beq}{\begin{eqnarray}}
\newcommand{\eeq}{\end{eqnarray}}
\begin{document}
\draft 
%\preprint{dvi file made on \today} 
 
\title 
{Deconstruction of the Trap Model for the New Conducting State in 2D}
\author{Philip Phillips}\vspace{.05in}

%
%\begin{instit}
\address
{Loomis Laboratory of Physics\\
University of Illinois at Urbana-Champaign\\
1100 W.Green St., Urbana, IL, 61801-3080}

%\end{instit}
\address{\mbox{ }}
\address{\parbox{14.5cm}{\rm \mbox{ }\mbox{ }
A key prediction of the trap model for the new conducting state in 2D is
that the resistivity turns upwards below some characteristic
temperature,
$T_{\rm min}$.  Altshuler, Maslov, and Pudalov\cite{alt} have
argued that the reason why no upturn has been observed
for the low density conducting samples is that the temperature
was not low enough in the experiments.
We show here that $T_{\rm min}$ within the Altshuler, Maslov, and
Pudalov
trap model
actually increases with decreasing density, contrary to their claim. 
Consequently,
the trap model is not consistent with the experimental trends. }}
\address{\mbox{ }}
\address{\mbox{ }}

%\columnseprule 0pt \narrowtext
\maketitle

In a series of recent papers, Altshuler, Maslov, and
Pudalov\cite{alt} (AMP)
proposed that the recent experimental finding by Kravchenko et al.,
Popovi\'{c} et al., Simmons et al., and Hanein et al.\cite{krav} 
of a new conducting state in a dilute 2D electron gas
is really
much ado about not very much. Namely, no new conducting state exists in
a dilute 2D electron gas, and all experiments\cite{review} observing a downturn in
the resistivity will eventually observe an upturn
at sufficiently low temperatures.  
In defense of this view, they offer\cite{alt} a trap model coupled with
arguments from weak localization in which
temperature-dependent traps are superimposed on a 
temperature-independent background potential.  Within this model they
predict that for a given concentration and strength of the trap
potential,
a downturn of the resistivity occurs but eventually the resistivity
turns around and increases at some characteristic temperature,
$T_{\rm min}$. They argue that $T_{min}$ should increase as the electron
density
increases. Consequently, saturation and eventual upturn of the
resistivity
should be easiest to observe in the high electron density samples.
 In fact,
such an upturn has been observed, thus far, only in
 the highest 
density samples\cite{sat,hamilton}, 
in apparent agreement with the prediction of the trap model.
While general criticisms\cite{krav2}
 have been levied at the
AMP\cite{alt} model, which
actually relies on four parameters to fit the experimental data, 
their calculation of $T_{\rm min}$ has not been addressed.
I show here that within the AMP model 
1) $T_{\rm min}$ in fact decreases 
with increasing electron density and
2) $T_{\rm min}$ is on the order of $1K$, both of which
are inconsistent with the experimental observations.  
Consequently, the lack 
of any upturn in the electrical resistivity in this temperature
regime
in the low electron density samples rules out the trapping model as a
viable interpretation of the experiments on the new conducting state.

Within a model that has both temperature-dependent and temperature
independent
disorder, AMP write the resistivity accordingly as
\beq
\rho_d(T)=\rho_1+\rho_0(T).
\eeq
In fact, a form of this type was first proposed by Pudalov for Si
samples
and later adopted
 in the context of the GaAs samples as a saturation of the resistivity
was observed at low temperatures. 
 Within the AMP
 model, the resistivity exhibits 
 a minimum at
\beq
T_{\rm min}=\frac{pa}{2}\frac{\rho_1^2}{d\rho_0/dT|_{T=T_{\rm min}}}
\eeq
where $p$ and $a$ are numerical constants.  In reaching the conclusion
that $T_{\rm min}$ increases with increasing electron density,
AMP used the experimental fact that the denominator, 
$d\rho_0/dT|_{T=T_{\rm min}}$, decreases
as the carrier density increases.  

It is unfortunate, however, that AMP did not consider the
density dependence of $\rho_1$, 
because to determine definitively the density dependence of a function, 
{\it both the denominator and
the numerator}, rather than only the denominator, 
must be considered.  The experiments clearly 
show that $\rho_1$, the resistivity from the residual scattering,
is strongly dependent on the carrier density, 
$n_s$.  For example, Hanein et al.\cite{han} have shown that 
$\rho_1$ is inversely proportional to $n_s-n_c$ in GaAs
heterostructures.   Nearly exponential density dependence
of $\rho_1^{-1}(n_s)$ was
also reported for Si in Ref.\cite{pud98} at $n_s\gtrsim n_c$. 
Inclusion of this effect leads to precisely the {\it opposite}
conclusion regarding the density dependence of $T_{\rm min}$.

To show this, I analyse the beautiful data of Pudalov et al. of Ref. 3b on
Si-MOSFET's.  Specifically, I focus on the data shown in the
inset of Fig. 3.  Shown there is a plot of the resistivity as a function
of temperature for 11 electron densities.  To consider the most
favourable case for the AMP scenario, I determined the slope of the
resistivity from its largest value. Because $T_{\rm 
min}$ is inversely proportional
to $d\rho_0/dT$, my estimate will then be a {\it lower bound}
to $T_{\rm min}$. Using the digitization feature
of Ghostview, I simply chose two points on the steepest part of 
$\rho(T)$ and then determined the slope. Consequently,
my analysis does not fall prey to the ambiguity suggested
in the response by AMP\cite{amp2}. In addition,  $\rho_1$ was obtained
from the extrapolated leveled value of $\rho(T)$ 
at zero tempearture.  I display in Figure 1 a plot of $T_{\rm min}$ 
versus the electron density obtained by analysing each
of the eleven curves shown in the inset of Figure 3 in Ref. 3b. Further,
to remove any ambiguity, I have provided the data
points used in the analysis in the figure caption.
As the figure clearly shows, $T_{\rm min}$ predicted by the AMP model
{\em decreases} (roughly as $1/n_s$), in contrast to their claim.
Hence, rather than corroborating the AMP scenario, the upturn
at high electron density now stands in stark
contrast to what their model actually predicts.  
\begin{figure}
\begin{center}
\epsfig{file=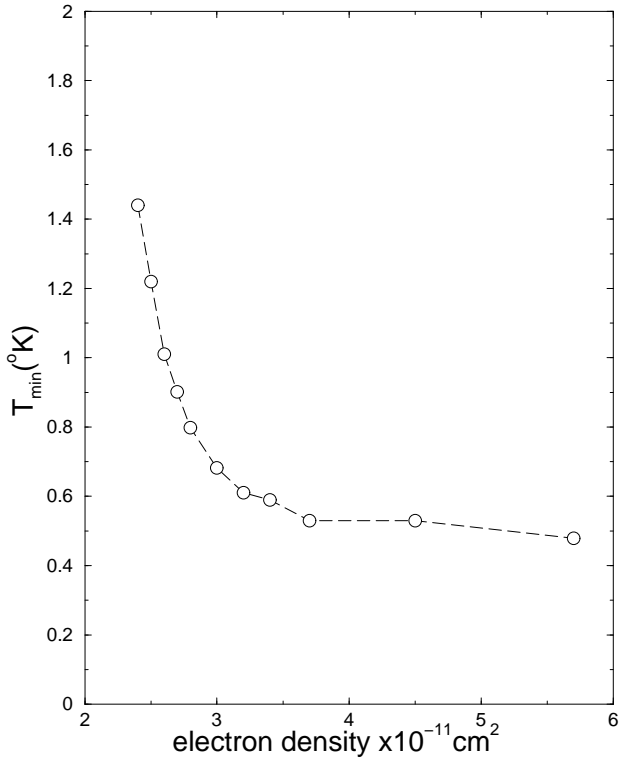, height=10cm}
\vspace{.2in}
\caption{Each
circle represents a calculation of 
$T_{\rm min}$ as determined by Eq. (2) as a function of 
electron density for each of the eleven Si-MOSFET 
samples reported in the
inset of Fig. 3 in Ref. 3b.  We see explicitly that $T_{\rm min}$
decreases as density increases in contrast to the claim of AMP.
For completeness, the data points correspond to the following 
($\rho_1,d\rho/dT$) pairs: (.4,.106),(.347,.094),(.294,.082),(.252,.0677),(.229,.063),(.188,.0496),
(.1529,.03686),(.129,.0271),(.1058,.02),(.076,.0086).}
\label{fig1}
\end{center}
\end{figure}

Further,
the $T_{\rm min}$'s determined here represent a lower bound
to the turnaround temperature. As these temperatures
are all on the order of 1K, they are certainly 
well accessible experimentally.  However, no such turnaround
has been observed in the experiments in the low density samples on
the conducting side.  
In fact, the recent finding
by Kravchenko and Klapwijk\cite{krav3} that the resistivity in
a low density
Si sample does not exhibit an upturn down
to 35 mK further points to the incorrectness of the
AMP model.

We close by pointing out that $\rho_0(T)\approx \exp(-T_0/T)$\cite{han}. 
Exponential decrease of the resistivity
is an indication that some sort of charge gap exists in the
 single particle spectrum.
Fermi liquids by definition cannot have a gap of any sort in the single particle spectrum.
In fact, no traditional metal has a charge gap in the single particle spectrum.
The only phase we know of that has a charge gap in the single particle spectrum
that conducts at zero temperature is a superconductor.  Hence,
this would suggest that experiments sensitive to pair formation
should be of utmost importance to the resolution of the nature of the charge
carriers in the new conducting
state in 2D.

 \acknowledgements
I thank Ladir Da Silva for help with the graphics and the DMR division
of the NSF for funding this work.  
%\end{multicols}
\end{document}